\documentclass[twocolumn]{iopart}
\usepackage{graphicx}

\begin{document}
\title{Mean-field approaches to the Bose-Hubbard model with three-body local interaction}
\author{Tomasz Sowi\'nski$^{1,2}$ and Ravindra W. Chhajlany$^{3,4}$}
\address{
$^1$ Institute of Physics of the Polish Academy of Sciences, Al. Lotnik\'ow 32/46, 02-668 Warszawa, Poland \\
$^2$ Center for Theoretical Physics of the Polish Academy of Sciences, Al. Lotnik\'ow 32/46, 02-668 Warszawa, Poland \\
$^3$ ICFO - Institut de Ci\`ences Fot\`oniques, Parc Mediterrani de la Tecnologia, E-08860
Castelldefels, Barcelona, Spain\\
$^4$ Faculty of Physics, Adam Mickiewicz University, 61-614 Pozna\'n, Poland
} 
\ead{Tomasz.Sowinski@ifpan.edu.pl}

\begin{abstract}
The zero temperature properties of the generalized Bose-Hubbard model including three-body interactions are studied  on a mean-field level. We obtain analytical results using the so-called perturbative mean-field method and more detailed numerical results using the Gutzwiller product state variational Ansatz. These two approaches yield equivalent results which compare well on a qualitative level with recent exact results obtained in the literature.
\end{abstract}
\maketitle

\section{Introduction}

The Superfluid (SF) to Mott-insulator (MI) transition of interacting bosonic particles  on a lattice is a paradigmatic  example of a quantum phase transition, introduced by Fisher \etal \cite{Fisher}. The realization in a seminal paper by Jaksch \etal \cite{Zoller1} that the Bose-Hubbard (BH) model can describe the dynamics of a system of ultra-cold atoms trapped in an optical lattice sparked interest in the experimental realization  of this model which culminated in the successful break-through experiment by the Munich group \cite{Greiner}. Rapid experimental progress in the experimental atomic and molecular physics community is leading to the successful simulation of various interacting models of both boson and fermion many-body systems using trapped atoms and molecules \cite{Zwerger,Lewenstein}. In step with these, various extensions of standard BH model have been vigorously studied theoretically \cite{Lahaye,Mering,SowinskiPRL,LewensteinBook}. A particularly interesting class of such  models take into account not only two- but also three-body local interactions.  
The phase properties  of these models have recently been widely discussed in the literature for one dimensional chains using exact diagonalization \cite{Sowinski2} and Density Matrix Renormalization Group methods \cite{Silva,Singh,Silva2,Sowinski3}. The so-called perturbative single site mean-field approach was used to obtain the qualitative phase boundaries in \cite{Chen,Zhou} for various parameter regimes. On the experimental side, effects of higher than two-body interactions have also been observed already \cite{Will}. Below, we consider the effect of three-body interactions at a mean-field level on repulsively interacting bosons undergoing the MI-to-SF quantum phase transition. 

\section{The Model}

The studied quantum many body system is described by the following Hamiltonian in grand canonical ensemble form:
\numparts  
\begin{eqnarray}
\hat{\cal H} &= \sum_i \hat{\cal U}_i -J \sum_{\{ij\}} \left(\hat{a}_i^\dagger \hat{a}_j +\hat{a}_j^\dagger \hat{a}_i\right) -\sum_i \mu\hat{n}_i, \label{Ham1} \\
\hat{\cal U}_i &= \frac{U}{2}\hat{n}_i(\hat{n}_i-1) + \frac{W}{6}\hat{n}_i(\hat{n}_i-1)(\hat{n}_i-2), \label{Ham2}
\end{eqnarray}
\label{Ham0}
\endnumparts   
where the sum $\sum_{\{ij\}}$ is over nearest neighbours in the optical lattice. The operator $\hat{a}_i$ annihilates a boson at site $i$ and $\hat{n}_i=\hat{a}_i^\dagger\hat{a}_i$  is the local density operator. The chemical potential $\mu$ controles the average number of particles in the system. The operator  ${\cal U}_i$ represents the on-site interaction between bosons. In contrast to the standard BH model where two-body interactions ($U$) are solely considered, the local interaction consists here also of three-body terms. The latter are represented in the last term in \eref{Ham2} and describe the energy cost of forming a local triple of bosons at a given lattice site given by the parameter $W$. 

In this Article, we study the stability properties of the zero temperature insulating phases in the system described by (1)  for repulsive two body interaction ($U>0$) within mean-field approaches. 
We recall here that the mean-field approach should be a good approximation for higher dimensional lattices, becoming exact in the limit of infinite lattice dimensionality. 
In this article, we compare results obtained by two mean-field approaches -- the perturbative mean-field and the Gutzwiller variational ansatz for the phase boundaries, and further use the latter approach to obtain the full phase diagram properties.

\section{Perturbative mean-field approach}
 
For the  considered problem, the superfluid state spontaneously breaking the $U(1)$ symmetry of the model (1) can be described by the local order parameter $\Phi_i = \langle \hat{a}_i \rangle$. In the mean-field approximation, one considers the annihilation operator to be described by its average value plus the operator describing fluctuations about this average $\hat{a}_i = \Phi_i + \hat{\delta}_i$ and neglecting second order fluctuations $\hat{\delta}_i\hat{\delta}_j$. In accordance with translational symmetry of the model (1), the order parameter is assumed uniform throughout the system and the hopping part of the Hamiltonian can be decoupled as $\hat{a}_i^\dagger \hat{a}_j = \Phi\hat{a}_i^\dagger + \Phi^*\hat{a}_j - |\Phi|^2$. In the studied case, without loss of generality, the order parameter can be set to be real. The resultant mean-field Hamiltonian  is a sum of local terms
\begin{equation}
\hat{\cal H}_{\mathtt{MF}} = \sum_i \hat{\cal U}_i - J z \Phi \left[\hat{a}_i^\dagger + \hat{a}_i - \Phi\right] - \mu\hat{n}_i,
\label{Hmf}
\end{equation}
where $z$ is coordination number of the lattice.  Due to the fact that in this description all sites are completely independent, one just considers a chosen site omitting the site index $i$. 

The order parameter should in principle be determined self-consistently from the condition $\Phi= \langle \hat{a} \rangle$ in the ground state of the single-site mean field Hamiltonian \eref{Hmf}. In particular, the boundary between the SF and MI insulator phases is determined by the vanishing of the SF order parameter (assuming the phase transition is continuous). The boundary can however be determined analytically by the following perturbative argument. The Mott phase (in mean-field description) corresponds to a Fock state corresponding to some integer filling. Moving across the quantum phase transition, the SF order parameter attains a small non-zero value and it contributes a perturbative term to the Hamiltonian of the system. Treating this perturbation to the lowest non-trivial order, one can ask when this term is energetically favorable, i.e. the SF ground state energy becomes lower than that of the Mott state, to obtain the phase boundary. 

For any $\mu$, the Mott state with integer filling $n_0$ has  ground-state energy ${\cal E}_0(n_0)={\cal U}(n_0)-\mu n_0$. To second order in $\Phi$, the perturbed ground-state energy is then given by
\begin{eqnarray}
{\cal E}(n_0) &= {\cal E}_0(n_0) + Jz\Phi^2 + \sum_{k\neq n_0}\frac{|\langle k|{\cal H}_{\mathtt{I}}|n_0\rangle|^2}{{\cal E}_0(n_0)-{\cal E}_0(k)} \nonumber \\
&={\cal E}_0(n_0) + Jz\Phi^2 \nonumber \\
&+ (Jz\Phi)^2\left[\frac{n_0+1}{{\cal E}_0(n_0)-{\cal E}_0(n_0+1)} + \frac{n_0}{{\cal E}_0(n_0)-{\cal E}_0(n_0-1)}\right],
\end{eqnarray}
where ${\cal H}_{\mathtt{I}} = -Jz\Phi(\hat{a}^\dagger + \hat{a})$ is the off-diagonal part of the on-site perturbation. To determine the phase boundary, it is convenient to rewrite the energy as
\begin{equation}
{\cal E}(n_0) = {\cal E}_0(n_0) + Jz\Phi^2 \left[1 - Jz{\cal F}(n_0) \right],
\end{equation}
where 
\begin{equation}
{\cal F}(n_0) = \frac{n_0+1}{{\cal E}_0(n_0+1)-{\cal E}_0(n_0)} + \frac{n_0}{{\cal E}_0(n_0-1)-{\cal E}_0(n_0)}.
\end{equation}
Now it is clear that whenever ${\cal F}(n_0)<(Jz)^{-1}$ the energy is minimized when the order parameter is zero and the system remains in the insulating phase. In contrast, for ${\cal F}(n_0)>(Jz)^{-1}$ the energy is minimized for non vanishing order parameter $\Phi$ and therefore the SF phase prevails. The phase boundary  is thus obtained from the limiting condition ${\cal F}(n_0)=(J_c z)^{-1}$ where $J_c$ is the critical value of tunneling. 

For the first insulating lobe, i.e. when $n_0=1$, the critical tunnelling expressed as a function of chemical potential $\mu$ is given by
\begin{equation}
J_c(\mu) = \frac{\mu(U-\mu)}{z(U+\mu)}
\end{equation}
and therefore, in particular,  the boundary is completely insensitive to three-body interactions. This observation is in contrast to results obtained with more accurate methods. For example, it was shown that, the tip of the Mott lobe shifts under strong three-body interactions in $d=1$ dimensions \cite{Silva,Sowinski2}. 

The second insulating lobe ($n_0=2$) is constrained by 
\begin{equation}
J_c(\mu)= \frac{(\mu-U)(2U+W-\mu)}{z(U+2W+\mu)}.
\end{equation}
In the limit of vanishing tunnelling $J\rightarrow 0$ the second insulating lobe is bounded by two values of chemical potential $\mu_-=U$ and $\mu_+=2U+W$. Moreover, its area defined as ${\cal S}=\int_{\mu_-}^{\mu_+} \mathrm{d}\mu\, J_c(\mu)$ scales as ${\cal S}\sim(U+W)^2$. Similar analysis can be extended to higher fillings in straightforward manner. The width of the insulating lobe for a given $n_0$ for zero tunnelling is equal to $U+(n_0-1)W$ and its area scales as ${\cal S}\sim[U+(n_0-1)W]^2$. Thus the three-body interactions have an increasingly important role for higher fillings in substantially stabilizing the Mott phase in comparison with the standard BH model.

\section{Gutzwiller ansatz}

In this section, the background necessary for obtaining the properties of the studied system  using the Gutzwiller variational ansatz is laid out \cite{Kivelson}. In order to capture both the Mott phase with fixed number of bosons per site, and the (mean field) SF phase which involves local boson number fluctuations, a product state ansatz over individual sites is assumed as an approximation to the true ground state:
$|\Psi\rangle = \prod_{i}|\psi_i\rangle$.
The local states $|\psi_i\rangle$ can be expanded in the  Fock basis, which for numerical purposes is truncated to some sufficiently large maximal number of particles $n_\mathtt{max}$. Thus the variational ansatz is of the form
\begin{equation}
|\Psi\rangle = \prod_{i}\sum_{n=0}^{n_\mathtt{max}}\alpha_i(n)|n\rangle_i,
\label{gutzwiller}
\end{equation}
where $\alpha_i(n)$ is the probability amplitude of finding $n$ bosons on site $i$. Note that on assuming translational invariance, $\forall_i |\psi_i\rangle = |\psi\rangle$, one is lead to the simple mean-field theory described in Section 2. The more general form of the product ansatz corresponds to the so-called unrestricted mean-field approximation. The basic advantage of the Gutzwiller ansatz, however, lies in the possibility of explicitly and directly obtaining the ground state instead of just chosen ground state expectation values. 

It is obvious that a product state can not posses any information about non-local correlations. Nevertheless, the SF long-range coherence is captured at this level  by non-trivial superpositions of local Fock states, i.e. the local order parameter 
$\langle \hat{a}_i\rangle$ is non-zero only when the local quantum state is a superposition of states with different occupations. 

The optimal state describing the state at a given parameter point of the studied Hamiltonian is determined variationally through energy minimization.  In practice the minimization procedure is implemented by using standard methods like the conjugate gradient method or imaginary-time evolution. The latter is based on the assumption that the dynamics of the system is determined by a mean-field Hamiltonian where the order parameter is not necessarily a priori uniform over all sites. The evolution in imaginary time in the infinite time limit then, in principle, projects an arbitrary initial state of the form \eref{gutzwiller} onto the ground state. The set of mean-field coupled equations of motion in imaginary time $\tau=it$ to be solved are 
\begin{eqnarray}
- \frac{d}{d \tau} \alpha_i(n) = &- J  \left[\sqrt{n}\,\alpha_i(n-1)  + \sqrt{n+1}\,\alpha_i(n+1)\right]\varphi_i\nonumber \\
& + \left[\frac{U}{2}n(n-1) + \frac{W}{6} n(n-1)(n-2) -\mu n \right] \alpha_i(n),
\end{eqnarray}
where $\varphi_i = \sum_{\{ij\}} \langle \hat{a}_j\rangle$ and the sum of the order parameters is over the sites coupled to site $i$ in the lattice. The $\varphi_i$ have to be determined self-consistently.

\section{Results}
\begin{figure}
\includegraphics{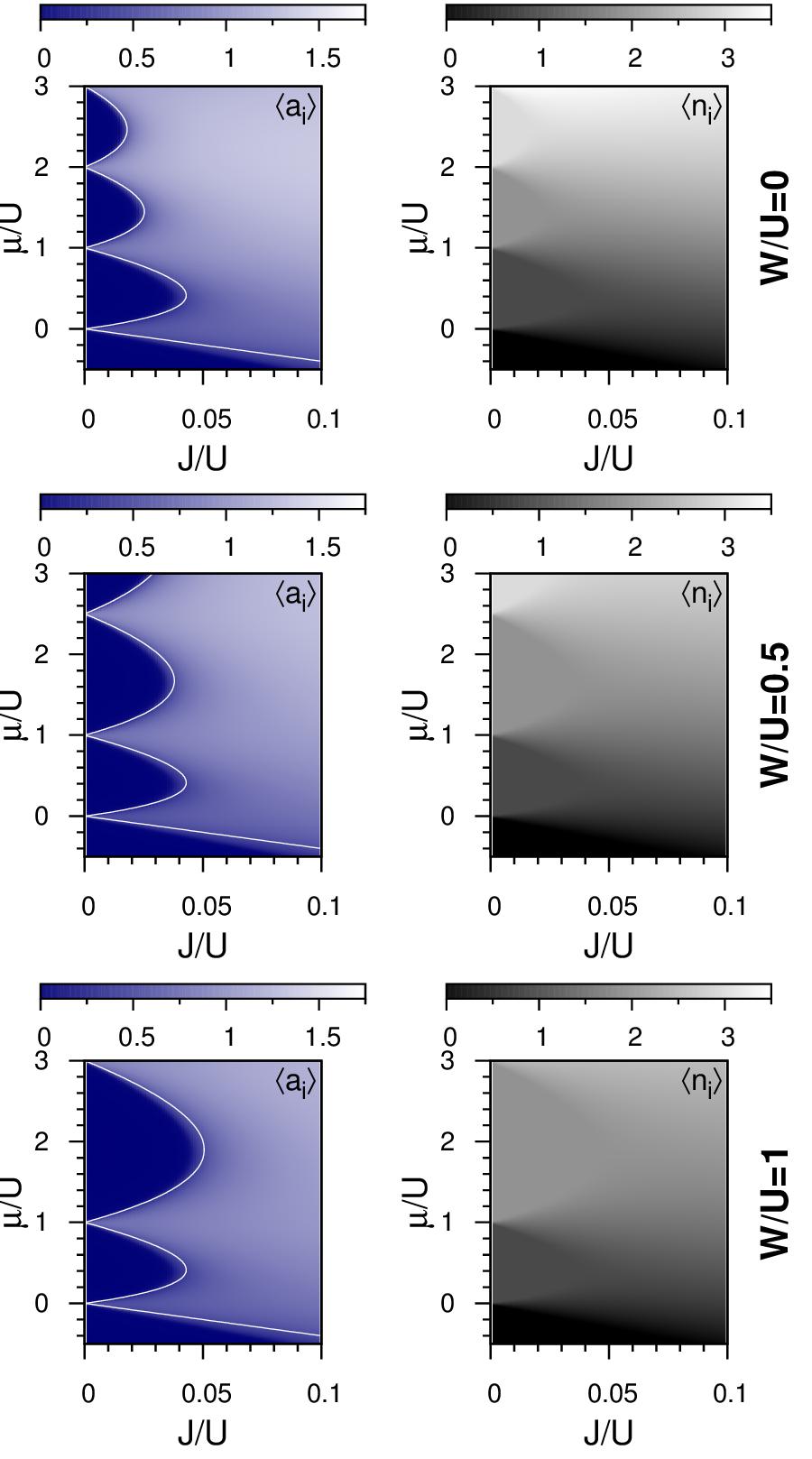}
\caption{Phase diagram of the system obtained using the Gutzwiller method for different values of three-body interaction strengths. The left panels depict the SF order parameters, while the right panels show the corresponding site occupations.  The phase boundary between MI and SF phases is clearly distinguished and in agreement with the perturbative mean-field analytical boundary (white line).  } \label{Fig1}
\end{figure}

Mean-field results are obtained for the studied system using the Gutzwiller method based on the imaginary time evolution technique and compared with the perturbative mean-field approach. The numerical calculations are performed on an $L\times L$ square lattice, with $L=8$ under periodic boundary conditions. The cut-off  $n_{\mathtt{max}}=4$. We note here that the Gutzwiller method does not assume translational invariance over the system and is in this sense  unbiased. 

Fig. 1 shows the phase diagrams obtained by the Gutzwiller method for  three values of 
three-body interaction strengths $W$. The ground state parametrization obtained using this method is translationally invariant as assumed in the perturbative mean field approach. The phase boundary is clearly identified by inspecting where the SF order parameter $\Phi$ becomes negligibly small which is also accompanied by commensurate filling of lattice sites. The phase boundary obtained by the perturbative mean-field is also plotted for comparison. The Gutzwiller results are generally in good agreement with the latter, although the Mott area around the lobe tips seem to be slightly underestimated using the number of simulation steps and cut-off chosen for the considered system size. 

As expected from perturbative mean field results, the Gutzwiller approach yields a diagram where the first lobe is unaffected by three-body interactions. Indeed, the particle superfluid formed above the upper half of the Mott lobe for unit filling has negligible contributions from states where two or more bosons occupy sites. The non-trivial effect of the three-body interaction is thus on the Mott lobes corresponding to higher lobes where three or more particles on site contribute to the ground state. On increasing (the repulsive) $W$, the primary effect with respect to the standard Bose-Hubbard model is the increase of Mott lobe widths and increased stability with respect to the hopping strength. Furthermore, the stability of the Mott lobe increases with increasing of the filling of the lattice. This is also in accordance with recent exact results obtained for the one-dimensional chain in the limit of dominant three-body interactions ($U=0$) \cite{Silva2,Sowinski3} where exactly the same effect was observed. Finally note that at the mean-field level, for general $W/U$, there is no change in character of MI-SF quantum phase transition which is still of second order. Again, this is in accordance with more exact results, e.g. in a one-dimensional studied using ED and DMRG showing no change of universality class, which there is the Kosterlitz-Thouless type \cite{Sowinski2,Sowinski3}. 

While the above discussion focussed on the repulsive case $W>0$, a similar simple analysis can be carried out for attractive interactions. There the lobes shrink in size with increasing filling number and for every value of $W$ there is a critical number of bosons above which translational invariance is lost and the bosons collapse on to one site resulting in a trivial phase separation on the lattice. 

\section{Acknowledgements}
This research was supported by the
(Polish) National Science Center Grants No. DEC-2011/01/D/ST2/02019 (T.S.) and DEC-2011/03/B/ST2/01903 (R.W.C.). R.W.C. acknowledges a Mobility Plus fellowship from the Polish Ministry of Science and Higher Education.


\section*{References}
\begin{thebibliography}{99}
\bibitem{Fisher} Fisher M P A \etal 1989 {\it Phys. Rev. B} {\bf 40} 546
\bibitem{Zoller1} Jaksch D \etal 1998 {\it Phys. Rev. Lett.} {\bf 81} 3108
\bibitem{Greiner} M. Greiner \etal 2002 {\it Nature (London)} {\bf 415} 39
\bibitem{Zwerger} Bloch I \etal 2008 {\it Rev. Mod. Phys.} {\bf 80} 885
\bibitem{Lewenstein} Lewenstein M \etal 2007 {\it Adv. Phys.} {\bf 56} 243
\bibitem{Lahaye} Lahaye T \etal 2009 {\it Rep. Prog. Phys.} {\bf 72} 126401
\bibitem{Mering} Mering A and Fleischhauer M 2011 {\it Phys. Rev. A} {\bf 83} 063630
\bibitem{SowinskiPRL} Sowi\'nski \etal {\it Phys. Rev. Lett.} {\bf 108} 115301
\bibitem{LewensteinBook} Lewenstein M \etal 2012 {\it Ultracold Atoms in Optical Lattices: Simulating quantum many-body systems} (Oxford University Press, Oxford)
\bibitem{Sowinski2} Sowi\'nski T 2012 {\it Phys. Rev. A} {\bf 85} 065601
\bibitem{Silva} Silva-Valencia J and Souza A 2011 {\it Phys. Rev. A} {\bf 84} 065601
\bibitem{Singh} Singh M \etal 2012 {\it Phys. Rev. A} {\bf 85} 051604(R)
\bibitem{Silva2} Silva-Valencia J and Souza A 2012 {\it Eur. Phys. J. B} {\bf 85} 161
\bibitem{Sowinski3} Sowi\'nski T 2013 {\it Preprint} arXiv:1307.6852
\bibitem{Chen} Chen B L. \etal 2008 {\it Phys. Rev. A} {\bf 78} 043603
\bibitem{Zhou} Zhou K \etal 2010 {\it Phys. Rev. A} {\bf 82} 013634
\bibitem{Will} Will S \etal 2010 {\it Nature (London)} {\bf 465} 197
\bibitem{Kivelson} Rokhsar D and Kivelson B 1991, {\it Phys. Rev. B} {\bf 44} 10328
\end{thebibliography}
\end{document}